\documentclass[preprint,12pt,3p]{elsarticle}
\usepackage{graphicx}% Include figure files
\usepackage{dcolumn}% Align table columns on decimal point
\usepackage[latin5]{inputenc}% For Turkish characters
\usepackage{amssymb,amsmath,mathptmx,amsbsy,bm}
\def\bra#1{\mathinner{\langle{#1}|}}
\def\ket#1{\mathinner{|{#1}\rangle}}

\journal{Physics Letters A}

\begin{document}

\begin{frontmatter}

\title{Deterministic transformations of bipartite pure states}

\author[label1]{Gokhan Torun}
\address[label1]{Department of Physics, Istanbul Technical University Maslak 34469, Istanbul, Turkey}
\ead{torung@itu.edu.tr}

\author[label1]{Ali Yildiz}
\ead{yildizali2@itu.edu.tr}

\begin{abstract}
We propose an explicit protocol for the deterministic transformations of bipartite pure states in any dimension using deterministic transformations in lower dimensions. As an example, explicit solutions for the deterministic transformations of $3\otimes 3$ pure states by a single measurement are obtained, and an explicit protocol for the deterministic transformations of $n\otimes n$ pure states by  three-outcome measurements is presented.
\end{abstract}


\end{frontmatter}

% \linenumbers


\section{Introduction}
Quantum entanglement  is
used as a resource in quantum information processes such as teleportation \cite{teleportation1}, dense coding \cite{densecoding} and quantum-key distribution \cite{key-dist}. The success probability in fulfilling quantum information tasks depends on the properties of the entangled state which is used as a resource \cite{agrawal,pati}. This requires a deep understanding of the transformation properties of entangled states under local operations and classical communication (LOCC). One line of research is the interconvertability  of multi-qubit states, and  the optimal \cite{Optimal_GHZ_distil,Yildiz,optimal1,optimal2,sheng1,deng,Torun} and deterministic \cite{turgut, tajima,zhao} transformations of some classes of multi-qubit states have been widely studied.  Another line of research is the interconversion of bipartite entangled states \cite{sheng2,sheng3}. In particular, Bennett et al. \cite{bennett} showed that the entanglement in any pure state of
a bipartite system can be concentrated by LOCC into maximally entangled states, and conversely, an arbitrary partly entangled state  of a bipartite system can be prepared by LOCC using maximally entangled states as the only source of entanglement. Vidal \cite{Vidal1} obtained the maximum transformation probability of $n\otimes n$ pure states in terms of the Schmidt coefficients and explicitly constructed a local protocol. Lo and Popescu \cite{Lo} showed that any general transformation between bipartite pure states using LOCC can be performed with
one-way classical communications only, and one way communication is more powerful than those without communications.  Chau et al. \cite{chau} presented  necessary and sufficient conditions for the probabilistic transformations of quantum states using local operations (without classical communication) only. Jonathan and Plenio \cite{jonathan} used a minimal set of entanglement monotones, and presented an optimal local strategy for entanglement concentration.

Deterministic transformation of a state by LOCC is of fundamental importance in quantum information theory, because if a state $\ket{\psi}$ can be transformed into another state $\ket{\phi}$ with unit probability by LOCC, then the information tasks that can be performed by using the state $\ket{\psi}$ can also be performed by using the state $\ket{\phi}$. He and Bergou \cite{He} showed that classical communication is necessary in realizing the deterministic transformations of a single bipartite entangled state.
Nielsen \cite{nielsen} used the algebraic theory of majorization and obtained the necessary and  sufficient condition for the deterministic transformations of bipartite pure states.
The same condition  was derived by using the method of areas  \cite{hardy}. Roa et al. \cite{roa} proposed a method for  the probabilistic transformation of  bipartite pure states based on local overlap modification and they obtained the deterministic transformation as a special case for states satisfying the majorization condition.
Majorization condition states that a state in Schmidt form
\begin{eqnarray}\label{111}
\ket{\psi}=\sum_{j=1}^{n}\psi_j \ket{j}\ket{j}, \quad (\psi_1 \geq \psi_2 \geq ... \geq \psi_{n} > 0)
\end{eqnarray}
can be transformed into another state
\begin{eqnarray}\label{222}
\ket{\phi}=\sum_{j=1}^{n} \phi_j \ket{j}\ket{j}, \quad (\phi_1 \geq \phi_2 \geq ... \geq \phi_{n} \geq 0)
\end{eqnarray}
by LOCC with unit probability if and only if $\lambda (\psi)$ is majorized by $\lambda (\phi)$, written $\lambda (\psi)\prec \lambda (\phi)$, where $\lambda (\psi)$ and $\lambda (\phi)$ denote the vectors of decreasingly ordered eigenvalues of the reduced density matrices $\rho_{A}^{\psi}$ and $\rho_{A}^{\phi}$, respectively. Majorization condition implies that the transformation $\ket{\psi}\rightarrow \ket{\phi}$ can be obtained with unit probability if and only if the inequality

\begin{eqnarray}\label{majorization}
\sum_{j=k}^{n}\phi_j^2 \leq \sum_{j=k}^{n}\psi_j^2
\end{eqnarray}
is satisfied for any $k$ $(1\leq k\leq n)$, with equality holding when $k=1$.

However, deterministic transformations of states by a single measurement  get more complicated as the dimension increases, since the construction of the doubly stochastic matrices, measurement operators and unitary operators gets more complicated \cite{nielsen-chuang}. Hence simple and explicit protocols  for the deterministic transformations of quantum states are of paramount importance. If the explicit solutions for the deterministic transformations of states in higher-dimensional space are known, then deterministic transformations of states in lower-dimensional space can be obtained as the special cases of higher-dimensional solutions.
However, the protocols for explicit transformations of states in higher-dimensional space using the lower-dimensional solutions still need to be developed.

In this Letter, we propose an explicit  protocol for deterministic transformations of $n\otimes n$  states using the solutions for the deterministic transformations of lower-dimensional quantum states $m\otimes m$ ($m<n$) . As an example, we find the explicit solutions for the deterministic transformations of $3\otimes 3$ pure states by a single three-outcome measurement followed by unitary transformations, and use them to construct a protocol for the deterministic transformation of $n\otimes n$  states in $\lfloor n/2 \rfloor$ steps.

\section{Deterministic transformations of $n\otimes n$ pure states}
In this section, we are going to present a protocol for the deterministic transformations of the source state $\ket{\psi}$ to the target state $\ket{\phi}$, satisfying the majorization condition, using the solutions for the deterministic transformations of lower-dimensional quantum states $m\otimes m$ ($m<n$). Suppose that the solutions for the deterministic transformation of
a state $\ket{\chi}=\sum_{j=1}^{m}\chi_j \ket{j}\ket{j}$ to any state $\ket{\omega}=\sum_{j=1}^{m}\omega_j \ket{j}\ket{j}$ satisfying the majorization condition, $\lambda (\chi)\prec \lambda (\omega)$, are known. Our protocol consists of   deterministic transformations of $m$-dimensional subspace of the source state in each step. We now describe the first step: The source state $\ket{\psi}$ can be written as
\begin{equation}
\ket{\psi}=\sum_{j=1}^{n-m}\psi_j \ket{j}\ket{j}+C\ket{\chi}
\end{equation}
where \begin{equation}
\ket{\chi}=\sum_{j=n-m+1}^{n}\chi_j \ket{j}\ket{j},\quad C=\sqrt{\sum_{j=n-m+1}^{n}\psi_j^2},\quad  \chi_j=\frac{\psi_j}{C}.
\end{equation}
The state $\ket{\psi}$ can be deterministically transformed to any state $\ket{\phi^{(1)}}$

\begin{equation}
\ket{\phi^{(1)}}=\sum_{j=1}^{n-m}\psi_j \ket{j}\ket{j}+C\ket{\omega},\quad \ket{\omega}=\sum_{j=n-m+1}^{n}\omega_j \ket{j}\ket{j}
\end{equation}
satisfying the majorization condition, $\lambda (\chi)\prec \lambda (\omega)$. We choose the state $\ket{\omega}$ such that maximum number of the Schmidt coefficients of the source state is transformed to the Schmidt coefficients of the target state. In the most general case, at least $m-1$ smallest Schmidt coefficients of the target state $\ket{\phi}$ can be obtained by choosing

\begin{equation}
\omega_j=\phi_j /C,\quad  n-m+2\leq j\leq n, \quad \omega_{n-m+1}=\sqrt{1-\sum_{j=n-m+2}^{n}\omega_j^2}.
\end{equation}
We note that only in the special case where the Schmidt coefficients of the source state $\ket{\psi}$ and target state $\ket{\phi}$ satisfy $\sum_{j=n-m+1}^{n}\psi_j^2=\sum_{j=n-m+1}^{n} \phi_j^2$, then $\omega_{n-m+1}=\phi_{n-m+1}/C$, and $m$ smallest  Schmidt coefficients of the source state can be transformed to $m$ smallest  Schmidt coefficients of the target state.

One may continue with $m$-dimensional deterministic transformations until the target state is obtained. The  transformation  $\ket{\psi}\rightarrow \ket{\phi}$ can be obtained  in $l$ steps

\begin{equation}\label{tf7}
 \ket{\psi}\rightarrow \ket{\phi^{(1)}}\rightarrow \ket{\phi^{(2)}}\rightarrow...\rightarrow \ket{\phi^{(l-1) }}\rightarrow \ket{\phi}
\end{equation}
where the intermediate states are given by
\begin{eqnarray}\label{fk7}
\ket{\phi^{(k)}}&=&\sum_{j=1}^{n-k(m-1)-1} \psi_j \ket{j}\ket{j}
+\tilde{\phi}_{n-k(m-1)}\ket{n-k(m-1)}\ket{n-k(m-1)} \nonumber \\
&& +\sum_{j=n-k(m-1)+1}^{n} \phi_j \ket{j}\ket{j},
\end{eqnarray}

\begin{equation}\label{n17}
\tilde{\phi}_{n-k(m-1)}=\sqrt{\psi_{n-k(m-1)}^2+\sum_{j=n-k(m-1)+1}^{n} \left(\psi_j^2-\phi_j^2\right)},\quad 1\leq k\leq l-1.
\end{equation}
By using eqs. \eqref{fk7}-\eqref{n17} it can be shown that the majorization conditions $\lambda (\phi^{(k)}) \prec \lambda (\phi^{(k+1)})$ are satisfied and hence the transformations $\ket{\phi^{(k)}}\rightarrow  \ket{\phi^{(k+1)}}$ can be obtained with unit probability.
In each step, at least $m-1$ more coefficients of the target state $\ket{\phi}$ are obtained, {\it i.e.}, the smallest $k(m-1)$ coefficients of $\ket{\phi^{(k)}}$ and $\ket{\phi}$ are equal. The states  $\ket{\phi^{(k)}}$ and $\ket{\phi^{(k+1)}}$ have $m$ nonequal Schmidt coefficients and the transformation $\ket{\phi^{(k)}}\rightarrow  \ket{\phi^{(k+1)}}$ is effectively an $m$-dimensional transformation. In the final transformation, $\ket{\phi^{(l-1)}}\rightarrow  \ket{\phi}$, the states    $\ket{\phi^{(l-1)}}$ and $\ket{\phi}$  may have $q$ ($2\leq q\leq m$) nonequal Schmidt coefficients which can be transformed by a single transformation.
In summary, using successive $m$-dimensional  deterministic transformations $l$ times, it is possible to obtain deterministic transformations of $n\otimes n$ ($(l-1)(m-1)+2 \leq n \leq (l-1)(m-1)+m$) bipartite pure states.

We now discuss the possibility of obtaining the target state, instead of transforming the smallest nonequal Schmidt coefficients, by transforming the greatest Schmidt coefficients in each step. In this case, the target state is again obtained, if possible, by successive deterministic transformations

\begin{equation}\label{tfm2}
\ket{\psi}\rightarrow \ket{\psi^{(1)}}\rightarrow \ket{\psi^{(2)}}\rightarrow...\rightarrow \ket{\psi^{(l-1)}}\rightarrow \ket{\phi}.
\end{equation}
In that case, the intermediate states turn out to be
\begin{eqnarray}\label{fk8}
\ket{\psi^{(k)}}=\sum_{j=1}^{k(m-1)} \phi_j \ket{j}\ket{j}
+\tilde{\psi}_{k(m-1)+1}\ket{k(m-1)+1}\ket{k(m-1)+1}
 +\sum_{j=k(m-1)+2}^{n} \psi_j \ket{j}\ket{j},
\end{eqnarray}

\begin{equation}\label{n18}
\tilde{\psi}_{k(m-1)+1}=\sqrt{\phi_{k(m-1)+1}^2+\sum_{j=k(m-1)+2}^{n} \left(\phi_j^2-\psi_j^2\right)},\quad 1\leq k\leq l-1.
\end{equation}
Using the majorization conditions it can be shown that $\tilde{\psi}_{k(m-1)+1}$ can be equal to zero, and  the number of nonzero Schmidt coefficients is reduced by one. In this case, the probability of obtaining the target state $\ket{\phi}$ from $\ket{\psi^{(k)}}$ will be zero, since the number of nonzero Schmidt coefficients cannot be increased by LOCC \cite{Lo}.
This rules out the possibility of any deterministic transformation other than the transformation of the smallest Schmidt coefficients in each step, as given by eq. \eqref{tf7}.

\section{Deterministic transformations of $3 \otimes 3$ pure states by a single measurement}

In this section, we are going to present a protocol for the deterministic transformation of the state

\begin{eqnarray}\label{3.32}
\ket{\psi}=a_1\ket{1}\ket{1}+b_1\ket{2}\ket{2}+c_1\ket{3}\ket{3},\quad (a_1\geq b_1 \geq c_1> 0)
\end{eqnarray}
to the state

\begin{eqnarray}\label{3.33}
 \ket{\phi}= a_2\ket{1}\ket{1}+b_2\ket{2}\ket{2}+c_2\ket{3}\ket{3},\quad (a_2\geq b_2 \geq c_2\geq 0)
\end{eqnarray}
satisfying the majorization condition by a single measurement. A generalized three-outcome measurement with the POVM elements leaving the Schmidt form invariant,
\begin{eqnarray}\label{op3}
M_i=\sqrt{\alpha_i}\ket{1}\bra{1}+\sqrt{\beta_i}\ket{2}\bra{2}+\sqrt{\gamma_i}\ket{3}\bra{3},\ \  \sum_{i=1}^{3}  M_i^\dag M_i =I,
\end{eqnarray}
is performed on one of the particles. The state after the measurement turns out to be one of the states

\begin{eqnarray}\label{3.2}
\ket{\psi_i}&=&\frac{M_i \ket{\psi}}{\sqrt{p_i}}
=\frac{1}{\sqrt{p_i}}\Big(\sqrt{\alpha_i} a_1\ket{1}\ket{1}+\sqrt{\beta_i} b_1\ket{2}\ket{2}+\sqrt{\gamma_i}c_1\ket{3}\ket{3}\Big)
\end{eqnarray}
with probabilities $p_i=a_1^2\alpha_i+b_1^2 \beta_i+c_1^2 \gamma_i$. We impose the condition that all three  states, $\ket{\psi_i}$, can be transformed to the state $\ket{\phi}$ by local unitary transformations, then the transformation $\ket{\psi}\rightarrow\ket{\phi}$ is a deterministic transformation. Although, due to the majorization condition, $a_1\leq a_2$ and $c_{1}\geq c_{2}$ there is not an unique relation between the other Schmidt coefficients, $\it {i. e.}$, both $b_1\geq b_2$ and $b_2\geq b_1$ are possible. We find that the local operations for the transformations turn out to be different for cases $b_1\geq b_2$ and $b_2\geq b_1$. We present the solutions for both cases:

\textit{$(i)$ The case  $b_1\geq b_2$;}
The three-outcome generalized measurement with POVM elements

\begin{eqnarray}\label{3.76}
M_1&=&\sqrt{p_1}\Big(\frac{a_2}{a_1}\ket{1}\bra{1}+\frac{b_2}{b_1}\ket{2}\bra{2}+\frac{c_2}{c_1}\ket{3}\bra{3}\Big), \nonumber \\
M_2&=&\sqrt{p_2}\Big(\frac{b_2}{a_1}\ket{1}\bra{1}+\frac{a_2}{b_1}\ket{2}\bra{2}+\frac{c_2}{c_1}\ket{3}\bra{3}\Big), \\ M_3&=&\sqrt{p_3}\Big(\frac{c_2}{a_1}\ket{1}\bra{1}+\frac{b_2}{b_1}\ket{2}\bra{2}+\frac{a_2}{c_1}\ket{3}\bra{3}\Big)\nonumber
\end{eqnarray}
transforms the state $\ket{\psi}$, given by eq. \eqref{3.32}, to one of the states
\begin{eqnarray}\label{3.73}
\ket{\psi_1}&=& a_2\ket{1}\ket{1}+b_2\ket{2}\ket{2}+c_2\ket{3}\ket{3},\nonumber \\
\ket{\psi_2}&=& b_2\ket{1}\ket{1}+a_2\ket{2}\ket{2}+c_2\ket{3}\ket{3},\\
\ket{\psi_3}&=& c_2\ket{1}\ket{1}+b_2\ket{2}\ket{2}+a_2\ket{3}\ket{3}\nonumber
\end{eqnarray}
with probabilities
\begin{equation}\label{3.81}
p_1=\frac{a_1^2}{a_2^2}-\frac{b_2^2}{a_2^2}\frac{(b_1^2 - b_2^2)}{(a_2^2 - b_2^2)}- \frac{c_2^2}{a_2^2}\frac{(c_1^2 - c_2^2)}{(a_2^2 - c_2^2)},\ \
p_2=\frac{b_1^2 - b_2^2}{a_2^2 - b_2^2},\ \
p_3=\frac{c_1^2 - c_2^2}{a_2^2 - c_2^2},
\end{equation}
respectively. The state $\ket{\psi_1}$ is already the target state $\ket{\phi}$, and the states $\ket{\psi_2}$ and  $\ket{\psi_3}$ can be transformed to the target state  by local unitary transformations $\ket{1}\leftrightarrow\ket{2}$ and $\ket{1}\leftrightarrow\ket{3}$, respectively. Since all states obtained after the measurement are transformed to the target state by local unitary transformations and total probability of success is unity ($p_1+p_2+p_3=1$), we may conclude that the deterministic transformation  $\ket{\psi}\rightarrow\ket{\phi}$ can be obtained by LOCC described above. The majorization condition and the condition  $b_1\geq b_2$ imply the ordering of the parameters $a_2\geq a_1\geq b_1\geq b_2 \geq c_2$, and hence for $a_2=b_2$ the source and target states are already equal.

\textit{$(ii)$ The case  $b_2\geq b_1$;}
The three-outcome generalized measurement with POVM elements

\begin{eqnarray}\label{3.77}
M_1&=&\sqrt{p_1}\Big(\frac{a_2}{a_1}\ket{1}\bra{1}+\frac{b_2}{b_1}\ket{2}\bra{2}+\frac{c_2}{c_1}\ket{3}\bra{3}\Big), \nonumber \\
M_2&=&\sqrt{p_2}\Big(\frac{c_2}{a_1}\ket{1}\bra{1}+\frac{b_2}{b_1}\ket{2}\bra{2}+\frac{a_2}{c_1}\ket{3}\bra{3}\Big), \\ M_3&=&\sqrt{p_3}\Big(\frac{a_2}{a_1}\ket{1}\bra{1}+\frac{c_2}{b_1}\ket{2}\bra{2}+\frac{b_2}{c_1}\ket{3}\bra{3}\Big)\nonumber
\end{eqnarray}
transforms the state $\ket{\psi}$ to one of the states
\begin{eqnarray}\label{3.74}
\ket{\psi_1}&=& a_2\ket{1}\ket{1}+b_2\ket{2}\ket{2}+c_2\ket{3}\ket{3},\nonumber \\
\ket{\psi_2}&=& c_2\ket{1}\ket{1}+b_2\ket{2}\ket{2}+a_2\ket{3}\ket{3},\\
\ket{\psi_3}&=& a_2\ket{1}\ket{1}+c_2\ket{2}\ket{2}+b_2\ket{3}\ket{3}\nonumber
\end{eqnarray}
with probabilities
\begin{equation}\label{3.82}
p_1=\frac{a_1^2}{a_2^2}-\frac{c_2^2}{a_2^2}\frac{(a_2^2 - a_1^2)}{(a_2^2 - c_2^2)}- \frac{(b_2^2 - b_1^2)}{(b_2^2 - c_2^2)},\ \
p_2=\frac{a_2^2 - a_1^2}{a_2^2 - c_2^2},\ \
p_3=\frac{b_2^2 - b_1^2}{b_2^2 - c_2^2},
\end{equation}
respectively. The state $\ket{\psi_1}$ is already the target state $\ket{\phi}$, and the states $\ket{\psi_2}$ and  $\ket{\psi_3}$ can be transformed to the target state  by local unitary transformations $\ket{1}\leftrightarrow\ket{3}$ and $\ket{2}\leftrightarrow\ket{3}$, respectively. Since all states obtained after the measurement are transformed to the target state by local unitary transformations and total probability of success is unity ($p_1+p_2+p_3=1$), we may conclude that the deterministic transformation  $\ket{\psi}\rightarrow\ket{\phi}$ can be obtained by LOCC described above. The majorization conditions, given by eq. \eqref{majorization}, and the condition  $b_2\geq b_1$ imply the ordering of the parameters $a_2\geq b_2\geq b_1\geq c_1 \geq c_2$, and hence for $b_2=c_2$, the source and target states are  equal.
We note that for $b_1=b_2$, the three-dimensional problem reduces to the two-dimensional problem, and both solutions, given by eqs. \eqref{3.76} and \eqref{3.77}, reduce to the solution of the two-dimensional problem.

Motivated by the solutions for the deterministic transformations of $3\otimes3$ states presented above, one may consider any possible generalization of the method to any dimension, \textit{i.e.}, deterministic transformation of \eqref{111} into \eqref{222} by a single measurement followed by local unitary transformations. The solution of the state transformation problem depends on the ordering of the Schmidt coefficients of the source and target states. There are only two different cases ($b_1\geq b_2$ and $b_2\geq b_1$) for transformations of $3\otimes3$ states as discussed above. In the $n\otimes n$-dimensional problem, there are at least $2^{n-2}$ inequivalent cases, each of which requires the solutions for highly nontrivial problem of finding local measurement operators and unitary transformations in high dimensions. The complexity of the problem increases exponentially as the dimension increases, and hence there is no simple generalization of deterministic state transformation by a single measurement to higher-dimensional spaces. However, it is possible to use the solutions obtained for transformation of states in lower dimensions to obtain transformation of states in higher dimensions.

\section{Deterministic transformations of  $ n\otimes n$ pure states by three-outcome measurements}
In this section, we are going to present a protocol for the deterministic transformations of the state given by eq. \eqref{111}, to the state given by eq. \eqref{222}, satisfying the majorization condition, using the results obtained for the deterministic transformations of $3\otimes 3$ pure states. Our protocol consists of $\lfloor n/2 \rfloor$ deterministic transformations,
\begin{equation}\label{tf}
 \ket{\psi}\rightarrow \ket{\phi^{(1)}}\rightarrow \ket{\phi^{(2)}}\rightarrow...\rightarrow \ket{\phi^{(\lfloor (n-2)/2 \rfloor)}}\rightarrow \ket{\phi},
\end{equation}
where the intermediate states are given by
\begin{equation}\label{fk}
\ket{\phi^{(k)}}=\sum_{j=1}^{n-2k-1} \psi_j \ket{j}\ket{j}+\tilde{\phi}_{n-2k}\ket{n-2k}\ket{n-2k}+\sum_{j=n-2k+1}^{n} \phi_j \ket{j}\ket{j},
\end{equation}

\begin{equation}\label{n1}
\tilde{\phi}_{n-2k}=\sqrt{\psi_{n-2k}^2+\sum_{j=n-2k+1}^{n} \left(\psi_j^2-\phi_j^2\right)},\quad 1\leq k\leq \lfloor (n-2)/2 \rfloor.
\end{equation}
In each step, two more coefficients of the target state $\ket{\phi}$ are obtained, {\it i.e.}, the smallest $2k$ coefficients of $\ket{\phi^{(k)}}$ and $\ket{\phi}$ are equal.

Deterministic transformations require  that the majorization conditions, $\lambda (\phi^{(k)}) \prec \lambda (\phi^{(k+1)})$, which lead to the inequalities given by

\begin{eqnarray}\label{c1}
\tilde{\phi}_{n-2k-2}^2+\phi_{n-2k-1}^2+\phi_{n-2k}^2 &\leq& \psi_{n-2k-2}^2+\psi_{n-2k-1}^2+\tilde{\phi}_{n-2k}^2 , \nonumber \\
\phi_{n-2k-1}^2+\phi_{n-2k}^2&\leq& \psi_{n-2k-1}^2+\tilde{\phi}_{n-2k}^2 , \\
\phi_{n-2k}&\leq& \tilde{\phi}_{n-2k} \nonumber
\end{eqnarray}
should be satisfied. By substituting $\tilde{\phi}_{n-2k}^2$, given by eq. \eqref{n1},  and using the majorization condition it can be shown that the inequalities given by eq. \eqref{c1} are satisfied.

We give the explicit solutions for the first transformation, $\ket{\psi}\rightarrow \ket{\phi^{(1)}}$, and the last transformation, $\ket{\phi^{(\lfloor (n-2)/2 \rfloor)}}\rightarrow \ket{\phi}$, for illustrative purposes, and other transformations can be done similarly.
For the first transformation, we write the states $\ket{\psi}$ and $\ket{\phi^{(1)}}$ as
\begin{eqnarray}\label{t1}
\ket{\psi}&=&\sum_{j=1}^{n-3}\psi_j \ket{j}\ket{j}+\mu \Big(a_1 \ket{n-2}\ket{n-2}+b_1 \ket{n-1}\ket{n-1}+c_1 \ket{n}\ket{n}\Big),\\
\label{t21}\ket{\phi^{(1)}}&=&\sum_{j=1}^{n-3}\psi_j \ket{j}\ket{j}+\mu \Big(a_2 \ket{n-2}\ket{n-2}+b_2 \ket{n-1}\ket{n-1}+c_2 \ket{n}\ket{n}\Big)
\end{eqnarray}
where
\begin{equation}\label{t13}
\mu=\sqrt{\psi_{n-2}^2+\psi_{n-1}^2+\psi_{n}^2},\quad a_1= \frac{\psi_{n-2}}{\mu}, \quad  b_1= \frac{\psi_{n-1}}{\mu}, \quad  c_1=\frac{\psi_{n}}{\mu},
\end{equation}
\begin{equation}
a_2=\frac{\tilde{\phi}_{n-2}}{\mu}= \sqrt{1-b_{2}^{2}-c_{2}^{2}}, \quad  b_2= \frac{\phi_{n-1}}{\mu}, \quad  c_2=\frac{\phi_{n}}{\mu}.\nonumber
\end{equation}
A three-outcome measurement with POVM elements

\begin{eqnarray}
M_{i}&=&\sum_{j=1}^{n-3} \sqrt{p_i}\ket{j}\bra{j}+\sqrt{\alpha_i}\ket{n-2}\bra{n-2}+\sqrt{\beta_i}\ket{n-1}\bra{n-1} \nonumber \\
&&+\sqrt{\gamma_i}\ket{n}\bra{n}, \quad p_i=a_1^2\alpha_i+b_1^2 \beta_i+c_1^2 \gamma_i,\quad i=1,2,3
\end{eqnarray}
should be performed on one of the particles.  The solutions for $\alpha_i$,  $\beta_i$, $\gamma_i$ and the local unitary transformations which should be done after the measurement depend on the parameters   $\phi_{n-1}$ and $\psi_{n-1}$.

\textit{$(i)$ The case  $\psi_{n-1}\geq \phi_{n-1}$}; A three-outcome measurement with POVM elements

\begin{eqnarray}
&M_1=\sqrt{p_1}\Big(\sum_{j=1}^{n-3}\ket{j}\bra{j}+\frac{a_2}{a_1}\ket{n-2}\bra{n-2}+\frac{b_2}{b_1}\ket{n-1}\bra{n-1}+\frac{c_2}{c_1}\ket{n}\bra{n}\Big),& \nonumber\\
&M_2=\sqrt{p_2}\Big(\sum_{j=1}^{n-3}\ket{j}\bra{j}+\frac{b_2}{a_1}\ket{n-2}\bra{n-2}+\frac{a_2}{b_1}\ket{n-1}\bra{n-1}+\frac{c_2}{c_1}\ket{n}\bra{n}\Big), & \\ &M_3=\sqrt{p_3}\Big(\sum_{j=1}^{n-3}\ket{j}\bra{j}+\frac{c_2}{a_1}\ket{n-2}\bra{n-2}+\frac{b_2}{b_1}\ket{n-1}\bra{n-1}+\frac{a_2}{c_1}\ket{n}\bra{n}\Big)&\nonumber
\end{eqnarray}
transforms the state $\ket{\psi}$, given by eq. \eqref{111}, to one of the states
\begin{eqnarray}
&\ket{\psi_1}= \sum_{j=1}^{n-3} \psi_j \ket{j}\ket{j}+\tilde{\phi}_{n-2}\ket{n-2}\ket{n-2}+\phi_{n-1}\ket{n-1}\ket{n-1}+\phi_{n}\ket{n}\ket{n},&\nonumber \\
&\ket{\psi_2}= \sum_{j=1}^{n-3} \psi_j \ket{j}\ket{j}+ \phi_{n-1}\ket{n-2}\ket{n-2}+\tilde{\phi}_{n-2}\ket{n-1}\ket{n-1}+\phi_{n}\ket{n}\ket{n},&\\
&\ket{\psi_3}= \sum_{j=1}^{n-3} \psi_j \ket{j}\ket{j}+\phi_{n}\ket{n-2}\ket{n-2}+\phi_{n-1}\ket{n-1}\ket{n-1}+\tilde{\phi}_{n-2}\ket{n}\ket{n}&\nonumber
\end{eqnarray}
with probabilities $p_1$, $p_2$ and $p_3$ given by eq. \eqref{3.81}.
The state $\ket{\psi_1}$ is already the state $\ket{\phi^{(1)}}$, and the states $\ket{\psi_2}$ and  $\ket{\psi_3}$ can be transformed to the  state $\ket{\phi^{(1)}}$ by local unitary transformations $\ket{n-2}\leftrightarrow\ket{n-1}$ and $\ket{n-2}\leftrightarrow\ket{n}$ respectively.

\textit{$(ii)$ The case  $\phi_{n-1}\geq \psi_{n-1}$};

\begin{eqnarray}
&M_1=\sqrt{p_1}\Big(\sum_{j=1}^{n-3}\ket{j}\bra{j}+\frac{a_2}{a_1}\ket{n-2}\bra{n-2}+\frac{b_2}{b_1}\ket{n-1}\bra{n-1}+\frac{c_2}{c_1}\ket{n}\bra{n}\Big),& \nonumber \\
&M_2=\sqrt{p_2}\Big(\sum_{j=1}^{n-3}\ket{j}\bra{j}+\frac{c_2}{a_1}\ket{n-2}\bra{n-2}+\frac{b_2}{b_1}\ket{n-1}\bra{n-1}+\frac{a_2}{c_1}\ket{n}\bra{n}\Big), & \\ &M_3=\sqrt{p_3}\Big(\sum_{j=1}^{n-3}\ket{j}\bra{j}+\frac{a_2}{a_1}\ket{n-2}\bra{n-2}+\frac{c_2}{b_1}\ket{n-1}\bra{n-1}+\frac{b_2}{c_1}\ket{n}\bra{n}\Big) &\nonumber
\end{eqnarray}
transforms the state $\ket{\psi}$ to one of the states
\begin{eqnarray}
&\ket{\psi_1}= \sum_{j=1}^{n-3} \psi_j \ket{j}\ket{j}+\tilde{\phi}_{n-2}\ket{n-2}\ket{n-2}+\phi_{n-1}\ket{n-1}\ket{n-1}+\phi_{n}\ket{n}\ket{n}, \nonumber \\
&\ket{\psi_2}= \sum_{j=1}^{n-3} \psi_j \ket{j}\ket{j}+ \phi_{n}\ket{n-2}\ket{n-2}+\phi_{n-1}\ket{n-1}\ket{n-1}+\tilde{\phi}_{n-2}\ket{n}\ket{n}, &\\
&\ket{\psi_3}= \sum_{j=1}^{n-3} \psi_j \ket{j}\ket{j}+\tilde{\phi}_{n-2}\ket{n-2}\ket{n-2}+\phi_{n}\ket{n-1}\ket{n-1}+\phi_{n-1}\ket{n}\ket{n}&\nonumber
\end{eqnarray}
with probabilities $p_1$, $p_2$ and $p_3$ given by eq. \eqref{3.82}.
The state $\ket{\psi_1}$ is already the state $\ket{\phi^{(1)}}$, and the states $\ket{\psi_2}$ and  $\ket{\psi_3}$ can be transformed to the state $\ket{\phi^{(1)}}$ by local unitary transformations $\ket{n-2}\leftrightarrow\ket{n}$ and $\ket{n-1}\leftrightarrow\ket{n}$ respectively.
The last transformation,  $\ket{\phi^{(\lfloor (n-2)/2 \rfloor)}}\rightarrow \ket{\phi}$, is effectively a two-dimensional transformation if $n$ is an even number, or  a three-dimensional transformation if $n$ is an odd number.

For $n$ is an even number, the states  $\ket{\phi^{(\lfloor (n-2)/2 \rfloor)}}$ and  $\ket{\phi}$ can be written as

\begin{eqnarray}\label{4.13}
\ket{\phi^{(\lfloor (n-2)/2 \rfloor)}}&=&\psi_1 \ket{1}\ket{1}+\tilde{\phi}_{2}\ket{2}\ket{2}+\sum_{j=3}^{n} \phi_j \ket{j}\ket{j},\nonumber \\
&=& \nu \left( a'_1 \ket{1}\ket{1}+b'_1 \ket{2}\ket{2}\right)+\sum_{j=3}^{n} \phi_j \ket{j}\ket{j},\\
\ket{\phi}&=& \nu \left( a'_2 \ket{1}\ket{1}+b'_2 \ket{2}\ket{2}\right)+\sum_{j=3}^{n} \phi_j \ket{j}\ket{j} \nonumber
\end{eqnarray}
where
\begin{equation}\label{t27}
a'_{1}= \frac{\psi_{1}}{\nu}, \quad  b'_{1}= \frac{\tilde{\phi}_{2}}{\nu} =\frac{\sqrt{\phi_{1}^2+\phi_{2}^2-\psi_{1}^2}}{\nu}, \quad
a'_{2}=\frac{\phi_{1}}{\nu}, \quad  b'_{2}= \frac{\phi_{2}}{\nu},\quad \nu=\sqrt{\phi_{1}^2+\phi_{2}^2}.
\end{equation}
A two-outcome measurement on one of the particles with the POVM elements
\begin{eqnarray}\begin{aligned}\label{4.18}
M'_1&=\sqrt{p'_1}\Big(\frac{a'_2}{a'_1}\ket{1}\bra{1}
+\frac{b'_2}{b'_1}\ket{2}\bra{2}+\sum_{j=3}^{n}\ket{j}\bra{j}\Big),\quad p'_1=\frac{(a'_{1})^2-(b'_2)^2}{(a'_2)^2-(b'_2)^2}, \\
M'_2&=\sqrt{p'_2}\Big(\frac{b'_2}{a'_1}\ket{1}\bra{1}
+\frac{a'_2}{b'_1}\ket{2}\bra{2}+\sum_{j=3}^{n}\ket{j}\bra{j}\Big), \quad p'_2=\frac{(a'_2)^2-(a'_1)^2}{(a'_2)^2-(b'_2)^2}
\end{aligned}\end{eqnarray}
can be used to obtain the deterministic transformation. If the measurement gives the outcome corresponding to the operator $M'_1$, with probability $p'_1$, then the resulting state is $\ket{\phi}$. But if the measurement gives the outcome corresponding to operator $M'_2$ , with probability $p'_2$, then both parties perform unitary operations $\ket{1}\leftrightarrow \ket{2}$ on their particles to obtain the state $\ket{\phi}$.

For $n$ is an odd number, the states  $\ket{\phi^{(\lfloor (n-2)/2 \rfloor)}}$ and  $\ket{\phi}$ can be written as

\begin{eqnarray}\label{4.111}
\ket{\phi^{(\lfloor (n-2)/2 \rfloor)}}&=&\psi_1 \ket{1}\ket{1}+\psi_2 \ket{2}\ket{2}+\tilde{\phi}_{3}\ket{3}\ket{3}+\sum_{j=4}^{n} \phi_j \ket{j}\ket{j},\nonumber \\
&=& \kappa \left( a''_1 \ket{1}\ket{1}+b''_1 \ket{2}\ket{2}+c''_1 \ket{3}\ket{3}\right)+\sum_{j=4}^{n} \phi_j \ket{j}\ket{j},\\
\ket{\phi}&=& \kappa \left( a''_2 \ket{1}\ket{1}+b''_2 \ket{2}\ket{2}+c''_2 \ket{3}\ket{3}\right)+\sum_{j=4}^{n} \phi_j \ket{j}\ket{j} \nonumber
\end{eqnarray}
where
\begin{equation}
a''_{1}= \frac{\psi_{1}}{\kappa}, \quad  b''_{1}=\frac{\psi_{2}}{\kappa},\quad c''_{1}= \frac{\tilde{\phi}_{3}}{\kappa}=\frac{\sqrt{\phi_{1}^2+\phi_{2}^2+\phi_{3}^2-\psi_{1}^2-\psi_{2}^2}}{\kappa},\nonumber
\end{equation}

\begin{equation}
a''_{2}= \frac{\phi_{1}}{\kappa},\quad b''_{2}= \frac{\phi_{2}}{\kappa},\quad c''_{2}= \frac{\phi_{3}}{\kappa},\quad \kappa=\sqrt{\phi_{1}^2+\phi_{2}^2+\phi_{3}^2}.
\end{equation}
The deterministic transformation of states, $\ket{\phi^{(\lfloor (n-2)/2 \rfloor)}}\rightarrow \ket{\phi}$, is effectively a three-dimensional transformation problem. The solutions for the three-outcome measurement of one the particles with POVM elements
\begin{eqnarray}\label{op4}
M''_i=\sqrt{\alpha''_i}\ket{1}\bra{1}+\sqrt{\beta''_i}\ket{2}\bra{2}+
\sqrt{\gamma''_i}\ket{3}\bra{3}+\sqrt{p''_i}\sum_{j=4}^{n}\ket{j}\bra{j},\ \  \sum_{i=1}^{3}  (M''_i)^\dag (M''_i) =I
\end{eqnarray}
and the unitary transformations can easily be found using the solutions for the three-dimensional case given by Eqs. \eqref{3.76}-\eqref{3.82}.

\section{Conclusion}
We have presented an explicit protocol for the deterministic transformation of any bipartite pure state of higher dimensions using the solutions of that of lower dimensions. We have divided the higher-dimensional transformation problem into some number of effectively smaller dimensional transformation problems. In each step, some number of Schmidt coefficients of the source state have been transformed to the  Schmidt coefficients of the target state. The constraint on each transformation is that the intermediate states obtained  by lower-dimensional transformations should be deterministically transformable to the target state, $\it {i. e.}$, each intermediate state should satisfy the necessary majorization conditions. This constraint also implies that  the number of nonzero Schmidt coefficients of the intermediate states cannot be less than that of the target state. We have shown that the transformation of greater nonequal Schmidt coefficients of the  source state could reduce the number of nonzero Schmidt coefficients of intermediate states, and obtaining the target state would be impossible. This leaves the transformations of the smallest nonequal Schmidt coefficients of the intermediate states to the smallest nonequal Schmidt coefficients of the target state as the only option. We have also shown that the intermediate states in our protocol satisfy the necessary majorization conditions.
As an example of the proposed protocol, we first obtain the deterministic transformations of $3 \otimes 3$ pure states which consist of a single three-outcome measurement, one-way classical communication and local unitary transformations. Then we use the results of  the deterministic transformations of $3 \otimes 3$ states, and construct the explicit operations for the deterministic transformations $n \otimes n$ states in $\lfloor n/2 \rfloor$ steps. We think that the proposed
protocol sheds some light on the transformations of pure states.

\section{Acknowledgments}
This work has been partially supported by the Scientific and
Technological Research Council of Turkey (TUBITAK) under Grant
113F256.

%% References


\bibliographystyle{elsarticle-num}

\end{document}